\documentclass[12pt,reqno]{amsart}
\usepackage[hypertex]{hyperref}
\usepackage[final]{graphicx}
\usepackage{amsfonts}
\usepackage{pdfsync}
\usepackage{amsmath}
\usepackage{amssymb}
\usepackage{amsthm}

\newtheorem{theorem}{Theorem}[section]
\newtheorem{lemma}[theorem]{Lemma}
\newtheorem{corollary}[theorem]{Corollary}

\newtheorem{definition}[theorem]{Definition}

\newcommand{\R}{{\mathord{\mathbb R}}}

\newcommand{\Sp}{{\mathord{\mathbb S}}}

\newcommand{\N}{{\mathord{\mathbb N}}}
\newcommand{\E}{{\mathord{\mathbb E}}}

\begin{document}
\title[Kinetic Theory]{Kinetic Theory and the Kac Master Equation}

\author[Carlen]{Eric Carlen $^1$}

\thanks{E.\ C.\ was supported in part by NSF grand DMS-0901632.}

\author[Carvalho] {Maria C. Carvalho $^2$}

\thanks{M. \ C. \ was supported by FCT grant  PTDC/MAT/100983/2008.}

\author[Loss]{Michael Loss$^3$}

\thanks{M.\ L.\ was supported in part by NSF grant DMS-0901304.}

\address{$^1$ Rutgers University , Department of Mathematics,, carlen@math.rutgers.ed}
\email{\href{mailto:carlen@math.rutgers.edu}{carlen@math.rutgers.edu}}

\address{$^2$ Department of Mathematics and CMAF, University of Lisbon, 1649-003 Lisbon, Portugal}
\email{\href{mailto:mcarvalh@cii.fc.ul.pt}{mcarvalh@cii.fc.ul.pt}}

\address{$^3$ Georgia Institute of Technology, School of Mathematics,
Atlanta, Georgia 30332-0160, loss@math.gatech.edu}
\email{\href{mailto:loss@math.gatech.edu}{loss@math.gatech.edu}}

\maketitle

\centerline{\large }

\vspace{.3truein}
\centerline{\bf Abstract}
This article reviews recent work on the Kac master equation and its low dimensional counterpart,
the Kac equation.  

\medskip {\sl .}

\section{Introduction}

\label{sec:introduction}

As a phenomenological equation, the Boltzmann equation is extremely successful. It is now synonymous with kinetic theory, i.e., the description of a large number of colliding particles.
For colliding hard spheres of radius $a$ in a box of volume $V$, it reads
\begin{eqnarray}\label{boltzmann}
& &{\partial \over \partial t}f + v \cdot \nabla_{x} f + F \cdot \nabla_{v}f  =\frac{a^2}{V^2} \int {\rm d} w \int_{S^2} {\rm d} e |(w-v)\cdot e|  \nonumber \\
&\times & \big [f(x, \frac{1}{2}[(v+w)+|v-w|e],t) f(x, \frac{1}{2}[(v+w)-|v-w|e],t) - f(x,v,t)f(x,w,t) \big] \nonumber
\end{eqnarray}
This equation together with the initial conditions $f(x,v,0) = f_0(x,v)$ is a non-linear initial value problem.
The function $f(x,v,t)$ is a probability density giving the probability $ f(x,v,t)dx dv$ for a particle to have its position and velocity in the cube $dx dv$ centered at $(x,v)$. In many ways, this type of equation can be used to describe any process that evolves under streaming and collisions, which is one of the reasons why this equation is so useful in many different circumstances.
Its foundation as the equation of kinetic theory is, however, mysterious. Up to this day, there is no satisfactory derivation of the Boltzmann equation starting from a classical Hamiltonian many body system, notwithstanding the work of Lanford \cite{MR0441164, MR0459449, MR0479206}. The problem is that the derivation works only up to the first average collision time. Even in this case, however, the derivation is a mathematical tour de force. Likewise, Illner and Pulvirenti \cite{MR985619,  MR849204}
consider the case of a gas jet entering a vacuum in the limit
where the number of particles tends to infinity but the particle density is small.  Since the gas is expanding in the vacuum
the number of collision among the molecules is finite. This allows them to obtain a result for all times.

An attractive feature of the Boltzmann equation is that it allows one to talk in precise terms what is meant by
approach to equilibrium. An equilibrium is  a stationary solution of the Boltzmann equation. In the absence of an external force $F$, it  is a straightforward computation to determine this function to be a global  Maxwellian distribution. There is the strong expectation that starting with any initial condition the solution should approach a global Maxwell distribution for large times.
 Considerable research has been done on proving return to equilibrium, i.e., studying the long--time behavior of this equation.  One of the main tools  in this endeavor, the $H$ functional or entropy, has been introduced by Boltzmann himself.  Boltzmann proved that $\int f \log f dx dv$ decreases with time.
A quantitative version of this result has been obtained by Desvilletes and Villani in \cite{MR2116276}. It is, however, a conditional theorem; their result holds for smooth global solutions of the Boltzmann equation. It is unknown whether such solutions exist.
The only result in this direction is the one of DiPerna-Lions
\cite{MR972541,MR1088276} . These global solutions, however, are weak and not amenable to the analysis in \cite{MR2116276}.

If one contemplates a bit the times scales between the validity of the Boltzmann equation proved up to now
and the desire to understand return to equilibrium, one may wonder what the connection of `return
to equilibrium' and microscopic physics might be. In any case it points to a huge gap in our understanding
of the long time evolution of many particle systems.

Since the Boltzmann equation is, for want of a better word, `paradigmatic' for describing systems consisting of
a large number of interacting agents in a probabilistic way, one might ask for a `derivation', not based on mechanical principles, but 
based on simple and transparent probabilistic assumptions.  This is the path taken by Kac in 1956 \cite{MR0084985}. 
A number of simplifications have to be made. The first is to restrict one's attention to  a spatially  homogeneous gas, i.e., where the distribution function does not depend on the position, i.e., one considers only the collision terms. This restriction is reasonable since one expects that collisions act on a different time scale than the streaming. 
In this case the Boltzmann equation takes the form, again written for hard spheres,
\begin{eqnarray}
& &{\partial \over \partial t}f  =  \frac{a^2}{V} \int {\rm d} w \int_{\Sp^2} {\rm d} e |(w-v)\cdot e|  \nonumber \\
&\times & \big [f( \frac{1}{2}[(v+w)+|v-w|e],t) f(\frac{1}{2}[(v+w)-|v-w|e],t) - f(v,t)f(w,t) \big]  \ .
\end{eqnarray}
Thus, one visualizes the times evolution of the full Boltzmann equation as a sequence of collisions interrupted by streaming.
This picture is used in the work of Desvilletes and Villani mentioned above and a large part of their work
goes into the analysis of the spatially homogeneous Boltzmann equation.
Indeed, it  presents substantial difficulties concerning the question of approach to equilibrium
and as a first step a further simplification can be made by considering Maxwellian molecules in which the collision
rate does not depend on the momentum transfer during the collision but just on the angles. The evolution equation
then takes the form
\begin{eqnarray} \label{threedkac}
& &{\partial \over \partial t}f  = \frac{1}{\tau} \int {\rm d} w \int_{\Sp^2} {\rm d} e B\left( \frac{(w-v)\cdot e}{|(w-v)|}\right)  \nonumber \\
&\times & \big [f( \frac{1}{2}[(v+w)+|v-w|e],t) f(\frac{1}{2}[(v+w)-|v-w|e],t) - f(v,t)f(w,t) \big] 
\end{eqnarray}
$$
\int_{\Sp^2} B\left(w\cdot e\right) {\rm d} e = \frac{1}{2} \int_{-1}^1 B(x) dx = 1 \ .
$$
The form of $B$ is not important but one should remark that for the case where the force law
is a repulsive $1/r^5$, the scattering is indeed independent of the momentum transfer and the function
$B$ can be expressed in terms of an elliptic function. This was discovered by Maxwell \cite{MR0044951}.
Kac simplified the model further by reducing the problem to one dimensional collisions. Two one dimensional colliding particles either go through each other without changing the momenta, or exchange the momenta. In order not to
limit the collision outcomes too much one drops the momentum conservation and only retains the energy conservation
and considers the Kac equation
\begin{equation} \label{onedkac}
{\partial \over \partial t}f  =  2 \int_{-\pi}^\pi {\rm d} \theta  \rho(\theta)\int_\R  {\rm d} w[f( v',t) f(w',t)- f(v,t)f(w,t)] 
\end{equation}
where 
\begin{equation}
v' = v \cos \theta  - w \sin \theta  \ , \  w' = v \sin \theta  + w\cos \theta   \nonumber
\end{equation}
are the `post collisional' velocities 
and where $\rho(\theta) {\rm d} \theta$ is a probability measure satisfying
\begin{equation}
\rho(-\theta) = \rho(\theta) \nonumber \ .
\end{equation}
This condition is known as local reversibility since the transitions $(v,w) \to (v', w')$ and $ (v',w') \to (v,w)$ are equally likely.
A particularly simple choice is
\begin{equation}
\rho(\theta)=\frac{1}{2\pi}  \nonumber \ .
\end{equation}
Among all the specializations done so far the really serious ones are the first two, namely passing to the spatially homogeneous case and restricting to Maxwellian Molecules.  Most of the results that hold for (\ref{onedkac})
can be extended to the (\ref{threedkac}).  This non--linear evolution equation can be derived from a master equation
as Kac has shown in \cite{MR0084985}.

The aim of this article is to describe some of the recent results concerning the connections between the
equation (\ref{onedkac}) and the Kac Master equation. Most of these results have been obtained in the last ten
years or so. If simple proofs are available they will be presented, otherwise the reader is referred to the literature.
In the next section we derive the  linear Kac master equation and formulate the question of approach to equilibrium in this new context. In the third section we the connection between the Kac Master equation and Kac's equation
(\ref{onedkac})
will be explained in terms of propagation of chaos. In the fourth section we calculate the gap for the master
equation and in the fifth section we talk about approach to equilibrium in the sense of entropy. 
We end the paper with a number of open problems.

\section{The Kac Master Equation}

Kac's starting point is a random walk in velocity space. Consider $N$ particles moving on the line.
As mentioned before, we shall restrict ourselves to the case where the gas is spatially homogeneous.
We denote by
$$
\vec v = (v_1, v_2,\cdots, v_N)
$$
the velocity vector, i.e., the coordinate $v_i$ is the velocity of the particle carrying the label $i$.
The total kinetic energy of this system is
$$
E= \frac{m}{2} \sum_{i=1}^N v_i^2 
$$
where we assume that all particles have the same mass $m$. We shall set $m=2$.
For the collision law we shall assume that 
$$
(v_i, v_j) \rightarrow ( v_i^{*}(\theta), v_j^*(\theta)) = (\cos \theta v_i -  \sin \theta v_j, \sin \theta v_i + \cos \theta v_j) 
$$
where the stared quantities are the post collisional velocities. Clearly, the kinetic energy is preserved.
Hence the state space of the system is described by points $\vec v$ on the sphere $\Sp^{N-1}(\sqrt E)$.
In physical collisions the momentum should be conserved too, however, this leads to exactly two outcomes:
Either the particles go through each other or exchange velocities. Such a collision process would not lead
to an ergodic Markov transition operator.

The ``Kac  walk''  can now be described as follows:

\medskip
\noindent
(1) Randomly pick a pair $(i,j)$ of distinct indices in $\{1, \dots, N\}$ uniformly among all such pairs. The particles with labels $i$ and $j$ will collide.

\medskip
\noindent
(2) Randomly pick a `scattering angle' $\theta$ with probability $\rho(\theta) d \theta$ on  $[0, 2\pi)$. 
 
\medskip
\noindent
(3) Update the velocities by a rotation, i.e.,
\begin{equation}\label{onedcollisionlaw}
(v_i, v_j) \rightarrow ( v_i^{*}(\theta), v_j^*(\theta)) = (\cos \theta v_i -  \sin \theta v_j, \sin \theta v_i + \cos \theta v_j) 
\end{equation}
Repeating this process yields a random walk on $\Sp^{N-1}(\sqrt E)$.

It is fairly straightforward to construct a Markov transition operator.
Let $\vec v_j$ be  the velocities of the particles after the $j$-th collision and let $\phi: \Sp^{N-1}(\sqrt E) \rightarrow \R$
be a continous function. Define the Markov transition operator $Q_N$ by
$$
Q_N\phi(\vec v) = \E\left\{ \phi(\vec v_{j+1} )| \vec v_j = \vec v\right\}
$$
where the expectation is with respect to the probabilities specified above, i.e.
\begin{equation} \label{qoperator}
Q_N \phi (\vec v) = \frac{1}{\left(N \atop 2\right)} \sum_{i < j} P_{i,j}\phi(\vec v)
\end{equation}
where
\begin{equation} \label{averagerotation}
P_{i,j} \phi(\vec v)=  \int_{-\pi}^\pi \rho(\theta)  \phi(R_{i,j,\theta} \vec v) {\rm d} \theta
\end{equation}
and
\begin{equation} \label{rotation}
R_{i,j,\theta} \vec v = (v_1, \dots ,v_i^*(\theta), \dots , v_j^*(\theta), \dots, v_N)  \ .
\end{equation}

\noindent
Let $F_0$ be the initial probability distribution for the velocities $\vec v$. The probability distribution $F_1$ of the velocities after one collision can be computed via
$$
\int_{\Sp^{N-1}(\sqrt E)}  \phi(\vec v) F_1(\vec v)  {\rm d} \sigma^{(N)}  = \int_{\Sp^{N-1}(\sqrt E)}\E\{ \phi(\vec v_1) | \vec v_0 =\vec v\} F_0(\vec v) {\rm d} \sigma^{(N)}
$$
i.e.,
\begin{equation} \label{weak}
\int_{\Sp^{N-1}(\sqrt E)}  \phi(\vec v) F_1(\vec v)  {\rm d} \sigma^{(N)}  = \int_{\Sp^{N-1}(\sqrt E)} Q_N \phi(\vec v) F_0(\vec v) {\rm d} \sigma^{(N)}
\end{equation}
Here $\sigma^{(N)}$ is the uniform normalized measure on $\Sp^{N-1}(\sqrt E)$. Since $\rho(-\theta) = \rho(\theta)$, the linear operator $Q_N$ is selfadjoint on $L^2( \Sp^{N-1}(\sqrt E), d \sigma^{(N)})$ and since (\ref{weak}) holds for all 
continuous functions $\phi$, we have that
$$
F_1(\vec v) = Q_NF_0(\vec v)  \ .
$$
Hence $F_j$ the probability distribution after $j$ collisions is given by
\begin{equation} \label{collisionmarkov}
F_j = Q_N^j F_0 \ .
\end{equation}

So far, time has not been mentioned, and here we need some further assumptions. It is reasonable to assume that the velocitiy distribution at time $t +dt$ should only depend on the distribution at time $t$ and not on what happened  in the past, i.e.,
we construct a continuous Markov process. Thus, we assume that between time $t$ and $t +dt$  
the probability that a collision occurs is 
$$
\lambda Q_N F(\vec v, t) dt + o(dt) \ ,
$$
where $\lambda$ is a constant. We assume that the probability that multiple collisions occur in the time interval $[t, t+dt]$ is negligible. 
Thus, since there are $N$ particles colliding independently, we have, for small time increments, 
$$
F(\vec v, t+dt) = N \lambda Q_N F(\vec v, t) dt + (1 - N \lambda F(\vec v, t)) dt +o(dt) \ ,
$$
or
$$
F(\vec v, t+dt) - F(\vec v, dt)  = N \lambda \left[ Q_N F(\vec v, t) dt   - F(\vec v, t)\right] dt +o(dt) \ .
$$
 Passing to the limit we find the {\bf Kac Master Equation}
$$
\frac{d}{dt} F(\vec v, t) = \lambda N \left[Q_N-I\right] F(\vec v, t) 
$$
$$
F(\vec v, 0)=F_0(\vec v) 
$$
where $F_0$ is the initial probability distribution.
 
\noindent
The initial value problem is now solved by the convergent power series
$$
F(\vec v, t) = e^{\lambda N(Q_N-I)t}F_0(\vec v) = \sum_{k=0}^\infty \frac{e^{-\lambda N t} (\lambda N t)^k}{k!} Q_N^k F_0 (\vec v) \ .
$$
Since $Q_N$ is an averaging operator, $Q_N 1 = 1$ and this implies that $F(\vec v, t)$ is a probability distribution
for all times $t$.  A simple computation shows that 
$$
\frac{1}{\left(N \atop 2\right)} \sum_{i < j}  \int_{-\pi}^\pi \rho(\theta)  \int_{\Sp^{N-1}(\sqrt E)} |F(R_{i,j, \theta} \vec v) -F(\vec v)|^2 {\rm d} \sigma^{(N)}
$$
$$
=2\left(\Vert F \Vert_2^2 -\langle F, Q_N F \rangle\right) \ ,
$$
where we denote the inner product on $L^2(\Sp^{N-1}(\sqrt E)$ by $\langle F, G \rangle$. 
It follows that $Q_N$ and hence $e^{\lambda N(Q_N-I)t}$ is ergodic, i.e., $Q_N F = F$ only if $F = 1$, in particular
$Q_N \le 1$.  
An immediate consequence of this is that $F(\cdot, t)$ converges to the function $1$ as time tends to infinity.
More precisely, a simple application of the spectral theorem leads to
\begin{lemma}
Assume that $0 \le F \in L^2(\Sp^{N-1}(\sqrt E), {\rm d} \sigma^{(N)})$ and that
$$
\int_{\Sp^{N-1}(\sqrt E) } F(\vec v) {\rm d} \sigma^{(N)} = 1 \ .
$$
Then we have {\bf approach to equilibrium} in $L^2$, i.e., 
$$
\Vert e^{\lambda N(Q_N-I)t} F_0 -1 \Vert_2 \to 0
$$
as $t \to \infty$. 
\end{lemma}
An issue we explore later is to find rates for the approach to equilibrium. For large systems it is generally observed that the rate of
equilibration is independent of the size of the system, i.e., $N$. Thus, the challenge is  to find  an estimate on the relaxation rate
that is uniform in the number of particles.

From now on we will adopt the following conventions.
{\it We shall henceforth assume that the function $F_0(\vec v)$ and hence $F(\vec v, t)$ is symmetric in the particle 
labels. We set $\lambda=1$. Finally, since the energy is extensive we can set $E=N$.}

With these choices, the Kac master equation reads
\begin{equation}
\frac{d}{dt} F (\vec v) = -\mathcal L F
\end{equation}
with the initial condition
\begin{equation}
F(\vec v, 0) = F_0(\vec v) \ ,
\end{equation}
where
\begin{equation}
\mathcal L =  N(I - Q_N ) \ ,
\end{equation}
and  $Q_N$ is given by (\ref{qoperator}), (\ref{averagerotation}) and (\ref{rotation}).
\section{Propagation of chaos}

Note that the velocities, viewed as random variables with respect to any probability measure on $\Sp^{N-1}$ are not independent since
$$
\sum_{j=1}^N v_j^2 = N \ .
$$
However, as $N$ gets large one would expect that this dependency gets weaker. The notion of chaos makes this precise.  It will be convenient to define it in the language of measure theory. To start we define the marginal measure. Let  $\mu^{(N)}$ be a probability measure on $\Sp^{N-1}(\sqrt N)$.  
Fix $k<N$  and pick any Borel $A \subset \R^k$. Define
$$
M_k(\mu^{(N)})[A] = \mu^{(N)}[\{(v_1, \dots, v_k) \in A\}] \ ,
$$
in other words, we integrate the measure $\mu^{(N)}$ over the sphere but restricting the variables $v_1, \dots , v_k$
to remain in the set $A$.
\vskip .1 true in
\noindent
\begin{definition}[Chaos]  Let $\mu$ be a given Borel probability measure on $\R$.  A sequence of probability measures
$\{ \mu^{(N)}\}_{N=2}^\infty$ is called $\mu$--{\bf chaotic}  if each $\mu^{(N)}$ is symmetric under permutations of the particle labels and 
for each positive integer $k$ the $k$--marginal measure $M_k(\mu^{(N)})$ converges to $\mu^{\otimes k}$, i.e., 
for every bounded, continuous test function $\chi(v_1, \dots, v_k)$ we have that
$$
\int \chi(v_1, \dots, v_k){\rm d} \mu^{(N)}(v_1, \dots, v_N) \rightarrow \int \chi(v_1, \dots, v_k) {\rm d}\mu(v_1) \cdots {\rm d} \mu(v_k)
$$
\end{definition}
An illuminating example is given by the Mehler  limit \cite{Mehler}. In this case the measure $\mu^{(N)}$ is given by the normalized
uniform surface measure which is symmetric. Further,
\begin{eqnarray}
& &\int \chi(v_1, \dots, v_k){\rm d} \sigma^{(N)}(v_1, \dots, v_N) \nonumber \\
&=& \frac{| \Sp^{N-1-k}(\sqrt N)|}{| \Sp^{N-1}(\sqrt N)|} \int \chi(v_1, \dots, v_k)
\left(1-\frac{\sum_{j=1}^k v_j^2}{N}\right)^{\frac{N-k-2}{2}} {\rm d}v_1 \cdots {\rm d}v_k \nonumber  \\
& \to& (2 \pi)^{-\frac{k}{2}}  \int \chi(v_1, \dots, v_k) e^{- \frac{\sum_{j=1}^k v_j^2}{2}} {\rm d}v_1 \cdots {\rm d}v_k \nonumber
\end{eqnarray}
\noindent
as $N \to \infty$.
For notational convenience we set
$$
\gamma^{(k)} = (2 \pi)^{-\frac{k}{2}} e^{- \frac{\sum_{j=1}^k v_j^2}{2}}  \ .
$$
Let us remark that Kac, in his 1956 paper talked about sequences that have the Boltzmann property, instead of chaotic sequences. In some ways this is a better name and the next theorem makes that clear. 
\begin{theorem}[Propagation of Chaos, Kac 1956]
Let $\{ F_0^{(N)} \sigma^{(N)}\}$ be a $f_0(v) {\rm d} v$--chaotic sequence and denote by $\{ F_t^{(N)} \sigma^{(N)}\}$ the sequence
of measures where  $F_t^{(N)}$ is the solution of the master equation, i.e.,  $F_t = e^{N(Q_N-I)t}F_0 $ for some fixed $t$. Then, 
$\{ F_t^{(N)} \sigma^{(N)}\}$ is a $f(v, t) {\rm d} v$--chaotic sequence and $f(v,t)$ is a solution of the initial value problem
\begin{equation}
{\partial \over \partial t}f  = 2 \int_{-\pi}^\pi {\rm d} \theta  \rho(\theta)\int_\R  {\rm d} w[f( v^*(\theta,t) f(w^*(\theta),t)- f(v,t)f(w,t)]  \nonumber
\end{equation}
with $f(v,0)=f_0(v)$, i.e., it is a solution of the Kac equation. Recall that $v^*(\theta)$ and $w^*(\theta)$ are given by
(\ref{onedcollisionlaw}).
\end{theorem}
 
For the proof we refer the reader to the original paper of Kac \cite{MR0084985}. 

The above theorem immediately raises the question whether any measure of the form $f(v)dv$ appears as the
marginal of a chaotic sequence. Or, more precisely, let  $f$ be a probability density on $\R$ with
$$
\int_\R vf(v) {\rm d} v = 0 \ , \ \int_\R v^2 f(v) {\rm d} v = 1 \ . 
$$
Is there an $f(v) {\rm d} v$ chaotic sequence $\{F^{(N)} \sigma^{(N)}\}$ ?  It is natural to consider $ f^{\otimes N}$ which
is a function on $\R^N$ and to restrict it to the sphere $\Sp^{N-1}(\sqrt N)$ and then to control the fluctuations of $\sum_{j=1}^N v_j^2$ with respect to the measure $ f^{\otimes N}{\rm d}v_1 \cdots {\rm d} v_N$.  Since the $v_i^2$ can be considered as independent random variables with
respect to $ f^{\otimes N}$, and since the fluctuations are controlled by the second moment of the random variables,
this amounts to an assumption on the fourth moment of the function $f(v)$. In fact we have the following theorem \cite{MR2580955}.
\begin{theorem}[CCLRV] \label{CCLRV}
Let $f$ be a probability density on $\R$ satisfying
$$
\int_\R f(v) v^2 {\rm d}v =1 \ , \ \int_\R f(v) v^4 {\rm d} v < \infty \ , f \in L^\infty(\R) 
$$
and let $\mu({\rm d}v) = f(v){\rm d} v$. Then $\{ [\mu^{\otimes N}]_{\Sp^{N-1}(\sqrt N)}\} $ is $\mu$ --chaotic.
\end{theorem}

\noindent
The core of the argument is the following central limit theorem. Set
$$
\Sigma := \sqrt{\int_\R (v^2-1)^2 f(v) {\rm d} v}  
$$
and
$$
Z_N(f,r):= \int_{\Sp^{N-1}(r)}\left( \frac{f}{\gamma}\right) ^{\otimes N} {\rm d} \sigma^{(N)}_r 
$$
\begin{theorem}\label{localcentrallimit}
Assume that 
$$
\int_\R vf(v) {\rm d} v = 0 \ , \ \int_\R v^2 f(v) {\rm d} v = 1 \ , \ \int_\R f(v) v^4 {\rm d} v < \infty \ , f \in L^p(\R) 
$$
for some $p>1$. Then
$$
Z_N(f,\sqrt N )= \frac{\sqrt 2}{\Sigma}(1+o(1))
$$
as $N \to \infty$.
\end{theorem}

\section{Kac's conjecture}

Recall that the solution of the Kac Master equation is formally given by
$e^{-\mathcal{L}_N t}$
where
$\mathcal{L}_N := N(I-Q_N)$. One measure of the rate of approach  to equilibrium is given by the gap of $\mathcal L_N$.
Define the gap
$$
\Delta_N = \inf \{ \langle F, \mathcal{L}_N F \rangle : F \perp 1 \ , \ \Vert F \Vert_2 =1 \} \ .
$$
If $f$ is any probability distribution that is in $L^2(\Sp^{N-1}(\sqrt N)$, 
it follows from the spectral theorem that
\begin{equation}
\Vert e^{-\mathcal{L}_N t}f - 1\Vert_2 \le e^{-\Delta_N t} \Vert f - 1\Vert_2 \ .
\end{equation}
The important question is
whether the gap persists as $N \to \infty$. In his 1956 paper Kac conjectured that
$$
\Delta_N \ge c >0
$$
where $c$ is independent on $N$. After some attempts in \cite{MR1743614} this conjecture was proved
 by E. Jeanvresse \cite{MR1825150} using H.-T. Yau's Martingale method.
An explicit expression for the gap was computed in \cite{MR1860691}. 
This result was rediscovered in \cite{MR1961868}.  The following theorem was proved in
\cite{MR1860691} (see also \cite{MR2020418} for more general results).
\begin{theorem}[CCL] Set $\rho(\theta) = \frac{1}{2\pi}$. We have that
$$
\Delta_N =  \frac{1}{2}\frac{N+2}{N-1}
$$
and the gap eigenfunction, unique up to a multiplicative constant, is given by
$$
F_{\Delta_N} = \sum_{j=1}^N \left (v_j^4 - \frac{3N}{(N+2)} \right)
$$
\end{theorem}
It is instructive to compare this result with the gap of the linearized Boltzmann equation.
The first marginal of $F_{\Delta_N}$ can be easily computed and one obtains
$$
\lim_{N \to \infty} M_1 F_{\Delta_N}(v) = (2 \pi)^{-1/2} e^{-\frac{v^2}{2}} (v^4 -6v^2 +3)  =  \gamma(v)  H_4(v)  \ .
$$
To linearize the Kac  operator, we set  $f = \gamma(v)(1+ \varepsilon h)$ and obtain
$$
\frac{1}{\pi} \int_{-\pi}^\pi {\rm d} \theta \int_{-\infty}^\infty {\rm d} w \gamma(w)\left[h(\cos \theta v - \sin \theta w)+h(\sin \theta v + \cos \theta w)
-h(v) - h(w)\right] 
$$
The fourth Hermite polynomial is an eigenfunction with eigenvalue $- \frac{1}{2}$ which is the gap of the linearized Kac operator.
In fact it is easy to see that all eigenfunctions are given by the Hermite polynomials.
(Mc Kean 66 \cite{MR0214112}, Gr\"unbaum 1972 \cite{MR0295718}). Thus, as $N \to \infty$ all the information one gets from the gap of the Master equation is the gap
of the {\it linearized} Boltzmann equation. It is likewise easy to see that all the eigenfunctions of the the operator $\mathcal L$ are given
by spherical harmonics. The tricky part, however, is to decide which among those is the gap eigenfunction.

The proof of the above theorem is not difficult and the method has been useful in a variety of circumstances
which gives us the reason to reproduce it here. In a first step one computes 
$$
\mathcal{L}_N F_{\Delta_N} = \frac{1}{2}\frac{N+2}{N-1} F_{\Delta_N} \ .
$$
Hence, $\Delta_N \le \frac{1}{2}\frac{N+2}{N-1}$. The real issue is to prove the reverse inequality.
We make an induction argument in the number of particles. For $N=2$
$$
\mathcal{L}_2 = 2(I-Q_2) 
$$
and $Q_2$ is a one dimensional projection. Hence
$$
\Delta_2 = 2 \ .
$$
 
\vskip .2 true in
\noindent
Write
$$
N(I - Q_N) =   \sum_{k=1}^N (I - Q^k_{N-1})
$$
where
$$
Q^k_{N-1}F = \frac{1}{\left(N-1 \atop 2\right)} \sum_{i < j, \ i, j \not= k} P_{i,j} F \ ,
$$
i.e., $Q^k$ is just the operator acting on the whole space with the interaction with particle $k$ being  absent.
Thus, for any $F \perp 1$ we have
$$
\langle F, N(I - Q_N) F \rangle =  \sum_{k=1}^N \langle F, (I- Q^k_{N-1} )F\rangle
$$
and we try to use the gap for $N-1$ particles, $\Delta_{N-1}$.
Note that for fixed $v_k$, the function $F(\cdot, v_k)$ is not perpendicular to the constant function on $\Sp^{N-2}(\sqrt{N-v_k^2})$. Thus, write
 $F = (F- P_kF)+P_kF $ where $P_kF(v_k)$ is the unique function with
$$
\int_{\Sp^{N-1}(\sqrt N)}  \phi(v_k) F(\vec v) {\rm d} \sigma^{(N)} = \int_{\Sp^{N-1}(\sqrt N)}  \phi(v_k) P_k F(v_k) {\rm d} \sigma^{(N)} 
$$
for all test functions $\phi(v_k)$, i.e.,
$$
P_kF(v_k) = \int_{\Sp^{N-2}(\sqrt{N-v_k^2})} F(\vec v) d \sigma^{(N-1) }
$$
$P_k$, as an operator on $L^2(\Sp^{N-1}(\sqrt N))$, is a selfadjoint projection.
Hence
\begin{eqnarray}
\langle F, N(I - Q_N) F \rangle &=& \sum_{k=1}^N \langle (F -P_kF) , (I- Q^k_{N-1}) (F-P_kF) \rangle \nonumber \\
\end{eqnarray}
since $Q^k_{N-1} P_kF=P_kF$.
By the induction assumption we know that
$$
\langle (F -P_kF) , (N-1) (I- Q^k_{N-1}) (F-P_kF) \rangle \ge  \Delta_{N-1} \Vert  (F -P_kF) \Vert^2  =\Delta_{N-1}[ \Vert F\Vert^2 -\langle F, P_k F \rangle] \ ,
$$
and hence
$$
\langle F, N(I - Q_N) F \rangle  \ge \frac{N}{N-1}  \Delta_{N-1}   [\Vert F \Vert^2 -\langle F, P F \rangle] 
$$
where 
$$
P = \frac{1}{N} \sum_{k=1}^N P_k \ .
$$
Setting
$$
\mu_N = \sup\{  \langle F, P F \rangle : F \perp 1 , \Vert F \Vert =1\}
$$ 
we find that
$$
\Delta_N \ge \frac{N}{N-1} \Delta_{N-1} (1-\mu_N) \ .
$$
Thus, we need to analyze the operator $P$. To this end we
set
$$
\pi_k(\vec v) = v_k \  \ \ , \ \ 
\frac{1}{N} \sum_{k=1}^N  P_k F = \mu_N F
$$
and notice that for $\mu_N \not= 0$, $F$ is a sum of functions depending only on one variable, i.e.,
if we set $P_kF = h \circ \pi_k$ 
$$
F = \frac{1}{\mu_N} \sum_{k=1}^N h \circ \pi_k
$$
Note that $h$ does not depend on $k$ since $F$ is symmetric.
Also note that
$$
\langle h\circ \pi_1, 1 \rangle = \langle h\circ \pi_1, (\pi_1)^2 \rangle = 0
$$
since
$$
\sum_{k=1}^N  (\pi_k)^2 = | \vec v|^2=N
$$
Now
$$
\frac{1}{N} \sum_{k=1}^N P_1  P_k F = \mu_N P_1F \ ,
$$
or
$$
\frac{1}{N} h\circ \pi_1 + \frac{N-1}{N} P_1  P_2 F = \mu_N  h\circ \pi_1 \ ,
$$
where
$$
P_1  P_2 F =: (Kh)\circ \pi_1
$$
Hence, if we denote by $\kappa_N$ the largest eigenvalue of $K$ subject to the conditions
$$
\langle h\circ \pi_1, 1 \rangle = \langle h\circ \pi_1, (\pi_1)^2 \rangle = 0
$$
then
$$
\mu_N = \frac{1}{N} + \frac{N-1}{N} \kappa_N
$$
and
$$
\Delta_N  \ge \frac{N}{N-1} \Delta_{N-1} (1-\mu_N) =  \Delta_{N-1} (1-\kappa_N)
$$
Thus, we have reduced the whole problem to the calculation of the spectrum of the
operator $K$, which is a relatively simple task. If $h$ is a function on the interval $[-\sqrt N, \sqrt N]$ then $h \circ \pi_2$
as a function on the sphere is constant on the circles of latitude perpendicular to the 2-direction. Now, averaging this function over all
rotations that fix the 1-direction delivers a new function which we denote by $(Kh) \circ \pi_1$. Thus, we have as a quadratic form
$$
\langle h\circ \pi_1 , (Kh) \circ \pi_1 \rangle = \langle  h\circ \pi_1 h\circ \pi_2 \rangle
$$
i.e., the $K$ --operator measures the correlation of $h\circ \pi_1$ and $h\circ \pi_2$
on the sphere. More generally, pick any two unit vectors $\vec e_1, \vec e_2$ and  consider the form
$$
 \langle  g(\vec v \cdot \vec e_1)  h(\vec v \cdot \vec e_2) \rangle = \langle   g(\vec v \cdot \vec e_1) (K_th)(\vec v \cdot \vec e_1) \rangle
$$
where $t =\vec e_1 \cdot \vec e_2$. A straightforward calculation yields
$$
(K_th)(v) = \frac{|\Sp^{N-3}|}{|\Sp^{N-2}|} \int_{-1}^1 h(vt +\sqrt{N-v^2}\sqrt{1-t^2} s) (1-s^2)^{\frac{N-4}{2}} {\rm d} s \ .
$$
In our situation $t=0$ and hence we obtain the explicit expression
$$
(Kh)(v) = \frac{|\Sp^{N-3}|}{|\Sp^{N-2}|} \int_{-1}^1 h(\sqrt{N-v^2} s) (1-s^2)^{\frac{N-4}{2}} {\rm d} s \ .
$$
The following lemma is proved in  \cite{MR1860691}.
\begin{lemma}
The eigenfunctions of $K$ are polynomials $p_n(v)$ with eigenvalues $\alpha_n$.
The eigenvalues vanish for $n$ odd and 
$$
\alpha_{2m} = (-1)^k \frac{|\Sp^{N-3}|}{|\Sp^{N-2}|} \int_0^\pi [\cos \theta]^{2m} [\sin \theta]^{N-3} {\rm d} \theta \ .
$$
In particular
$$
\alpha_2 = -\frac{1}{N-1} \ , \alpha_4 = \frac{3}{N^2-1}
$$
$|\alpha_{2m}|$ is decreasing in $m$.
\end{lemma} 

\vskip .2 true in
\noindent
The second largest eigenvalue of $P$ is therefore
$$
\Delta_N   \ge \Delta_{N-1} \left[ 1-  \frac{3}{N^2-1} \right]
$$
Since
$\Delta_2 = 2$
we easily find
$$
\Delta_N \ge \frac{1}{2}\frac{N+2}{N-1}  \ ,
$$
which proves the theorem.

It is somewhat gratifying that the gap can be computed in other circumstances too.
An interesting particular case is where the distribution of the scattering angle is not uniform
and given by the density  $\rho(\theta)$. It can be easily verified that
$F_{\Delta_N}$ is always an eigenfunction
of $\mathcal{L}_N$ with eigenvalue
$$
\Gamma_N :=\frac{1}{4} \frac{(N+2)}{N-1} \Gamma_2
$$
where
$$
\Gamma_2 = 2  \int_{-\pi}^\pi (1 -\cos(4\theta)) \rho(\theta) {\rm d} \theta \ .
$$
However, $\Gamma_N$ is not the gap in general, since for $N=2$ we may have
$$
\Delta_2 = 2 \min_{k \ge 1} \int_{-\pi}^\pi (1-\cos (k\theta)) \rho(\theta) {\rm d} \theta < \Gamma_2
$$
Our induction scheme applied naively, leads to
$$
\Delta_N \ge \frac{\Delta_2}{4} \frac{N+2}{N-1}
$$
which proves Kac's conjecture in this case too.
We can do better, however.
\begin{theorem}[CCL]
Assume that
$$
\Delta_2 > 0.45 \Gamma_2 
$$ 
then for all $N$ sufficiently large $\Delta_N =\Gamma_N$ and $F_{\Delta_N}$ 
is the corresponding eigenfunction.
\end{theorem}
For the proof, which is a bit trickier, we refer the reader to the paper \cite{MR2020418}

The Kac Master equation for a spatially homogeneous gas of particles in three dimensional space is analogous to the one dimensional case. Recall that the collision law between two particles is given by
\begin{eqnarray}
& & v_i^*(w) = \frac{1}{2}[(v_i+v_j)+|v_i-v_j|w]\nonumber \\
& & v_j^*(w) =\frac{1}{2}[(v_i+v_j)-|v_i-v_j|w] \nonumber \ ,
\end{eqnarray}
where $w \in \Sp^2$.
These collision preserves now the energy sphere and the momentum plane, i.e., the quantities
$$
\sum_{j=1}^N |v_j|^2  \ , \ {\rm and} \ \sum_{j=1}^N v_j 
$$
are preserved. As before, we fix the total energy to be $N$ and, in addition, we fix the total momentum to be zero.
The Kac operator is obtained by replacing $P_{i,j}$ in $Q_N$ by
$$
P_{i,j} F = \int_{\Sp^2} F(v_1, \dots, v^*_i(w), \dots, v^*_j(w), \dots, v_N) B\left(\frac{(v_i-v_j)}{|v_i-v_j|}\cdot w\right) {\rm d} w  \ .
$$
It is a selfadjoint operator on the space $L^2(M, \mu_N)$ where $M$ is the intersection of the energy
sphere with the momentum plane. The measure $\mu_N$ is the Euclidean measure on $\R^{3N}$ restricted
to $M$.

It was shown \cite{MR2020418} for the case where $B$ is constant, that $\Delta_N \ge c >0$, $c$ independent of $N$.
We can, however, say much more. Set
$$
B_1 = \frac{1}{2} \int_{-1}^1 xB(x) {\rm d} x \ \ {\rm and} \ \ B_2 = \frac{1}{2} \int_{-1}^1 x^2 B(x) {\rm d} x
$$
and recall that $\frac{1}{2} \int_{-1}^1 B(x) {\rm d} x = 1$. The following two theorems were proved in \cite{MR2403324}
\begin{theorem}[CGL] 
Suppose that $B_2 > B_1$ and that 
$$
\Delta_2 \ge \frac{20}{9}(1-B_2) \ .
$$
Then for all $N \ge 3$ 
$$
\Delta_N = \frac{N}{N-1} (1-B_2) \ .
$$
Moreover, the eigenspace is three dimensional, and is spanned by the functions
$$
\Phi(\vec v) = \sum_{j=1}^N |v_j|^2 v_j^\alpha \ , \alpha=1,2,3
$$
\end{theorem}
 \begin{theorem}[CGL] Suppose that $\Delta_2 =2(1-B_1)$. Then for all $N \ge 7$,
 $$
 \Delta_N = \min \left \{(1-B_1), \frac{N}{N-1} (1-B_2)\right\} \ .
 $$
 Moreover, if $B_2 > B_1$, the eigenspace is three dimensional, and is spanned by the functions
 $$
 \Phi(\vec v) = \sum_{j=1}^N |v_j|^2 v_j^\alpha  \ , \ \alpha=1,2,3
 $$
 If $B_2 < B_1$, the eigenspace is spanned by the functions of the form
 $$
 |v_i|^2-|v_j|^2 \ \ {\rm and} \ \ v_i^\alpha -v_j^\alpha \ , \ \alpha =1,2,3
 $$
 for all $i < j$.
  \end{theorem}
The overall strategy of the proof is the same as in the case for one--dimensional collisions, i.e.,
the problem is reduced to the study of a low dimensional operator $K$. The eigenvalues, however, do not have
any obvious monotonicity properties and the computations are much more complicated.  Detailed estimates on Jacobi polynomials are used. We refer the interested reader to  \cite{MR2403324}.

The computations of the gaps for the various models can be considered a success, the drawback is that one learns
about as much about the approach to equilibrium as through the linearized Boltzmann equation. More importantly,
the notion of gap has other drawbacks and this will be explained in the next section.
 
\section{Entropic approach to equilibrium}
Using the gap as a rate for approach to equilibrium has an obvious drawback.
Assume that $F_0 = \prod_{j=1}^N f_j$ subject to the normalization condition
$$
\int_{\Sp^{N-1}(\sqrt N)} F_0(\vec v)  {\rm d}\sigma^{(N)} = 1 \ .
$$
The almost independence of the functions $f_j$ yield that
$$
\Vert F_0(\cdot)\Vert_2 \approx  \prod_{j-1}^N \int_{\Sp^{N-1}(\sqrt N)}  f_j^2 \rm d \sigma^{(N)}  = e^{{\rm const.} N} \ .
$$
Hence, the same is true for $\Vert F_0-1\Vert$ and, using the gap estimate only, it will take a time of order $N$
to relax to the equilibrium distribution.
The right quantity to consider is Boltzmann's relative {\it entropy} 
$$
H(F |\sigma^{(N)})  := \int_{\Sp^{N-1}(\sqrt N)}  F(\vec v) \log F(\vec v)  {\rm d} \sigma^{(N)}\ .
$$
 In general
if $\mu, \nu$ are two probability measures, their relative entropy is defined by
$$
H(\mu|\nu) = \int h \log h d\nu \ , \ \ h =\frac{{\rm d} \mu}{{\rm d} \nu} \ .
$$
Thus, if  $f(v) {\rm d} v$ is a probability measure on $\R$ then the relative entropy of $f \rm d v$ with respect to the Gaussian function $\gamma(v)$  is given by
$$
H(f|\gamma) = \int_\R f(v) \log \frac{f(v)}{\gamma(v)} dv
$$
Note that if $F_N$ is an $f(v) {\rm d}v$ chaotic family one would expect that
$$
H(F_N |\sigma^{(N)})  = N  H(f|\gamma)
$$
as $N \to \infty$. In other words, the entropy is, like the total energy, an extensive quantity, i.e., proportional to $N$.
It has been shown by Boltzmann that $H(f|\gamma)$ decreases in time for solutions of the Boltzmann equation.
This is the famous H-Theorem. For the Kac equation, this can be readily seen since
\begin{eqnarray}
& & \frac{\rm d}{\rm d t} H(f(\cdot, t)|\gamma) = \frac{\rm d}{\rm d t} \int_\R f(v,t) \log f(v,t) {\rm d} v 
-  \frac{\rm d}{\rm d t}  \int_\R f(v,t) \log \gamma(v) {\rm d} v \nonumber \\
&=& 2\int_{-\pi}^\pi {\rm d} \theta \rho(\theta) \int_\R \int_\R \left[f(v',t)f(w',t) - f(v,t)f(w,t)\right] \log f(v,t) {\rm d} v {\rm d} w \nonumber 
\end{eqnarray}
\begin{eqnarray}
&= &\int_{-\pi}^\pi {\rm d} \theta \rho(\theta) \int_\R \int_\R \left[f(v',t)f(w',t) - f(v,t)f(w,t)\right] \log[ f(v,t)f(w,t)] {\rm d} v {\rm d} w  \nonumber \\
&=& - \int_{-\pi}^\pi {\rm d} \theta \rho(\theta) \int_\R \int_\R \left[f(v',t)f(w',t) - f(v,t)f(w,t)\right]  \nonumber \\
&\times&  \left[\log f(v',t)f(w',t) -  \log f(v,t)f(w,t)\right] {\rm d} v {\rm d}w  \le 0\nonumber
\end{eqnarray}
\noindent
Note that we have used the fact that the second moment, the kinetic energy, is preserved in time to drop the term
$$
\frac{\rm d}{\rm d t}  \int_\R f(v,t) \log \gamma(v) {\rm d} v \ .
$$
This raises immediately the question for the rates of equilibration \cite{MR1964379}. Is there an exponential rate?
One could ask the same question for the Kac master equation, i.e.,
is it true that 
$$
\frac{\rm d}{\rm d t} H(F(\cdot, t)  |\sigma^{(N)}) \le -{\rm const.} H(F(\cdot, t)  |\sigma^{(N)}) \ ,
$$
for a constant that is, hopefully, independent of $N$. This surmise for the Boltzmann equation is known as Cercigniani's conjecture. 
The best result so far is by Cedric Villani \cite{MR1964379} who proved 
\begin{theorem}[Villani's Theorem]
Let $F_0$ be any probability density on $\Sp^{N-1}(\sqrt N)$ with finite relative entropy $H(F_0(\cdot)  |\sigma^{(N)}) $.
Then the solution of Kac's master equation with initial condition $F_0$ satisfies
$$
H(F(\cdot, t)  |\sigma^{(N)}) \le e^{-C_N t} H(F_0(\cdot)  |\sigma^{(N)}) 
$$
where
$$
C_N = \frac{2}{N-1} \ .
$$
\end{theorem}

Differentiating $H(F(\cdot, t))  |\sigma^{(N)}) $ with respect to time yields
$$
\frac{d}{dt} H(F(\cdot, t)  |\sigma^{(N)}) = - \int_{\Sp^{N-1}(\sqrt N)}  \left[\mathcal{L}_NF(\vec v, t)\right]  \log F(\vec v, t)  {\rm d} \sigma^{(N)}
$$
The term
$$
\int_{\Sp^{N-1}(\sqrt N)}  \left[\mathcal{L}_NF(\vec v)\right]  \log F(\vec v)  {\rm d} \sigma^{(N)}
$$
is called the entropy production.  Villani proved that for all densities $F$, 
\begin{eqnarray}
& &\int_{\Sp^{N-1}(\sqrt N)}  \left[N(I-Q_N)F(\vec v)\right]  \log F(\vec v)  {\rm d} \sigma^{(N)} \nonumber \\
& \ge & \frac{2}{N-1} \int_{\Sp^{N-1}(\sqrt N)} F(\vec v)  \log F(\vec v)  {\rm d} \sigma^{(N)} \nonumber
\end{eqnarray}
that is, the relative entropy production is bounded by $2/(N-1)$.
\begin{proof}[Sketch of a proof]
The idea is to interpolate  densities via the heat kernel on the sphere. Recall that
$$
\Delta =  \sum_{i<j} L_{i,j}^2
$$
where $L_{i,j}$ is an angular momentum operator, i.e.,
$$
L_{i,j}= v_i \partial_j - v_j \partial_i \ .
$$
Set
$$
F(s)= e^{\Delta s} F
$$
and compute
$$
\frac{\rm d}{\rm d s} \int F(s) \log F(s) {\rm d} \sigma^{(N)} = \int \Delta F(s) \log F(s) {\rm d} \sigma^{(N)}
$$
$$
=-\int \frac{| \nabla F(s)|^2}{F(s)} {\rm d} \sigma^{(N)} = -4 \int | \nabla \sqrt{F(s)} |^2{\rm d} \sigma^{(N)} \ .
$$
Define $\Delta^{k,l}$ by
$$
\Delta = \sum_{i<j} L_{i,j}^2 = \Delta^{k,l} +L_{k,l}^2 \ .
$$
Since
$$
\left[L_{i,j}, \Delta\right] = 0 \ {\rm and} \ 
\left[ \Delta^{k,l}, L_{k,l}\right] = 0
$$
we have that
$$
e^{\Delta s}  = e^{\Delta^{k,l}  s} e^{L_{k,l}^2s }
$$
and hence
$$
e^{\Delta s} F(x)  = e^{\Delta^{k,l}  s} [e^{L_{k,l}^2  s} F](x)  =\int G^{k,l} (x, y; s)  [e^{L_{k,l}^2  s} F](y) {\rm d} \sigma^{(N)}(y) \ ,
$$
where $G^{k,l} (x, y; s)$ is the kernel associated with the heat semigroup $ e^{\Delta^{k,l}  s} $. It has the properties
$$
G^{k,l}(x,y,s) \ge 0 \ , \ \int G^{k,l} (x, y;s)  {\rm d} \sigma^{(N)} (y)= 1 \ .
$$

Set 
$$
F_{k,l} (s)= e^{L_{k,l}^2  s} F
$$
and note that
$$
\lim_{s \to \infty} F_{k,l} (s) = P_{k,l}F \ .
$$
Since 
$$
-4 \int | \nabla \sqrt{F(s)} |^2{\rm d} \sigma^{(N)}  = -4 \sum_{k<l}  \int | L_{k,l} \sqrt{F(s)} |^2{\rm d} \sigma^{(N)}  \ .
$$
and the function
$$
F \rightarrow  | L_{k,l} \sqrt{F(s)} |^2
$$
is convex,  we can use Jensen's inequality to find the lower bound
$$
  -4 \sum_{k<l}\int \int G^{k,l} (x, y;s)  | L_{k,l}  \sqrt{F_{k,l}(y, s)} |^2{\rm d} \sigma^{(N)}(y)  {\rm d} \sigma^{(N)}(x) 
$$
$$
=-4 \sum_{k<l} \int \int G^{k,l} (x, y;s) {\rm d} \sigma^{(N)}(x) | L_{k,l} \sqrt{F_{k,l}(y, s)} |^2{\rm d} \sigma^{(N)}(y)
$$
$$
= -4  \sum_{k<l} \int  | L_{k,l} \sqrt{F_{k,l}(y, s)} |^2{\rm d} \sigma^{(N)}(y) \ .
$$
To summarize, we have shown that
$$-4 \int | \nabla \sqrt{F(s)} |^2{\rm d} \sigma^{(N)}  \ge -4\sum_{k<l}  \int  | L_{k,l}\sqrt{F_{k,l}(y, s)} |^2{\rm d} \sigma^{(N)}(y)
$$
or
$$
\frac{d}{ds} \int F(s) \log F(s) {\rm d} \sigma^{(N)}  \ge -4\ \sum_{k<l}  \int  | L_{k,l}\sqrt{F_{k,l}(y, s)} |^2{\rm d} \sigma^{(N)}(y)
$$
$$
= -  \sum_{k<l} \int  \frac{|L_{k,l} F_{k,l}(y, s) |^2}{ F_{k,l}(y, s)}{\rm d} \sigma^{(N)}(y)
$$
$$
= \sum_{k<l} \frac{d}{ds} \int  F_{k,l}\log F_{k,l} {\rm d}\sigma^{(N)} \ .
$$
Integrating both sides yields
$$
- \int F \log F {\rm d} \sigma^{(N)}  \ge \sum_{k<l}  \left[ \int P_{k,l} F \log P_{k,l}F{\rm d}\sigma^{(N)}  - \int F \log F {\rm d} \sigma^{(N)}\right]
$$
or
$$
\left[\left(N \atop 2\right) -1\right]\int F \log F {\rm d} \sigma^{(N)}   \ge  \sum_{k<l}    \int P_{k,l} F \log P_{k,l}F{\rm d}\sigma^{(N)}
$$
$$
\ge \sum_{k<l}    \int P_{k,l} F \log F {\rm d}\sigma^{(N)} =\left(N \atop 2\right) \int Q_NF \log F {\rm d}\sigma^{(N)}
$$
or
$$
 \int (I-Q_N) F \log F {\rm d}\sigma^{(N)} \ge \frac{1}{\left(N \atop 2\right)} \int F \log F {\rm d} \sigma^{(N)}
 $$
 from which Villani's theorem follows.
\end{proof} 
Villani's theorem yields an equilibration time that is again of the order $N$, since we have for all densities $F$
\begin{equation} \label{entropyproduction}
\frac{ \langle  \left[\mathcal{L}_NF \right] ,  \log F \rangle}
{H(F, \sigma^{(N)})}  \ge \frac{2}{N-1} \ .
\end{equation}
This raises the obvious question whether the estimate can be  improved.
Is there a density  $F_N$ so that 
$$
\lim_{N \to \infty}\frac{ \langle  \left[\mathcal{L}_NF_N\right] ,  \log F_N \rangle}
{H(F_N, \sigma^{(N)})} = 0 \ .
$$
The following theorem is proved in \cite{MR2580955}.
\begin{theorem}[CCLRV] \label{CCLRVtwo}
 For each $c>0$, there is a probability density $f$ on $\R$ with 
$$
\int_\R vf(v){\rm d} v = 0 \ , \ \int_\R v^2 f(v) {\rm d}v =1 \ ,
$$
an $f(v){\rm d} v$ chaotic family $\{F_N\}_{N \in \N}$ such that 
$$
\limsup_{N \to \infty} \frac{ \langle  \left[\mathcal{L}_NF_N(\vec v, t)\right] ,  \log F_N(\vec v, t) \rangle}
{H(F_N, \sigma^{(N)})} \le c \ .
$$
For each $c$ the density is smooth, bounded and has moments of all order.
\end{theorem}
If one considers the Kac equation ( \ref{onedkac}) instead of the master equation, then the intuition behind this theorem is not difficult to understand. Denote by $\gamma_a(v)$ the normalized Gaussian on $\R$
centered at the origin of variance $a$. Now consider the function
$$
f_\delta(v) = \delta \gamma_{\frac{1}{2\delta}}(v) + (1-\delta) \gamma_{\frac{1}{2(1-\delta)}}(v)
$$
where $\delta$ is a small number. The first  Gaussian describes an ensemble of a {\it small} fraction of particles
that have almost all of the kinetic energy, whereas the remaining large fraction of the particles have very little energy.
Intuitively, one expects that it would take a long time for such a state to equilibriate. In fact, it is not hard to see that
that the entropy production of this state is of order $-\delta \log \delta$, and thus small. 
With this function $f_\delta(v)$ one can construct a $f_\delta$ - chaotic state $F_N$ in the spirit of Theorem \ref{CCLRV} which has the properties stated in the previous theorem. Note that, as $\delta$ shrinks to zero, the fourth moment tends to infinity.
Villani \cite{MR1964379} conjectured that there exists a constant $C$ such that for every $N$ there is $F_N$ with
$$
\lim_{N \to \infty}\frac{ \langle  \left[\mathcal{L}_NF_N\right] ,  \log F_N \rangle}
{H(F_N, \sigma^{(N)})}  <\frac{C}{N} \ .
$$

This conjecture was essentially proved recently by Amit Einav \cite{Einav}.
\begin{theorem}
\label{cor:tnropy production result}For any $0<\beta<\frac{1}{6}$
there exists a constant $C_{\beta}$ depending only on $\beta$ such
that\[
\Gamma_{N}\leq\frac{C_{\beta}\log N}{N^{1-2\beta}}\]
\end{theorem}
The crux of the matter is to make the statement in Theorem \ref{CCLRVtwo}, the statement about a chaotic state and its marginal quantitative.The key theorem
is an analog of Theorem \ref{localcentrallimit}, which can be rendered in a much stronger form since it is
a statement about the particular function $f_\delta(v)$.
\begin{theorem}
\label{thm:approximation of Z}Let $f_{\delta_{N}}(v)=f_{\delta_{N}}(v)=\delta \gamma_{\frac{1}{2\delta_{N}}}(v)+(1-\delta_{N})\gamma_{\frac{1}{2(1-\delta_{N})}}(v)$
where $\delta_N$ is chosen such that
\begin{equation}
\delta_{N}^{1+2\beta}\cdot N\underset{N\rightarrow\infty}{\longrightarrow}\infty \ ,
\delta_{N}^{1+3\beta}\cdot N\underset{N\rightarrow\infty}{\longrightarrow}0 \ . \end{equation} 
Then for a fixed $j$ \[
Z_{N-j}\left(f_{\delta_{N}},\sqrt{u}\right)=\frac{2}{\sqrt{N-j}\cdot\Sigma_{\delta_{N}}\cdot|\mathbb{S}^{N-j-1}|u^{\frac{N-j}{2}-1}}\left(\frac{e^{-\frac{\left(u-N+j\right)^{2}}{2(N-j)\Sigma_{\delta_{N}}^{2}}}}{\sqrt{2\pi}}+\lambda_{j}(N-j,u)\right)\]
where $\sup_{u\in\mathbb{R}}\left|\lambda_{j}(N-j,u)\right|\leq\epsilon_{j}(N)$
and $\lim_{N\rightarrow\infty}\epsilon_{j}(N)=0$. 
\end{theorem}

Theorem \ref{cor:tnropy production result} or rather its proof is an application of the interplay between
the Kac equation and the Kac Master equation.  In this particular instance knowing a state of low entropy production for the Kac equation yields a state of low entropy production for the Kac Master equation.
This connection can be cast as entropic chaos.
\noindent
Let $\mu$ be a probability measure on $\R$ and for each $N$ let $\mu^{(N)}$ be a probability measure on $S^{N-1}(\sqrt N)$.
The sequence $\{ \mu^{(N)}\}_{N \in \N} $ is said to be {\bf entropically $\mu$--chaotic} if it is $\mu$--chaotic and in addition
$$
\lim_{N \to \infty} \frac{H(\mu^{(N)}|\sigma^N)}{N} = H(\mu|\gamma)
$$ 
\begin{theorem}
Let $f$ be a probability density on $\R$ satisfying
$$
\int_\R f(v) v^2 {\rm d} v =1 \ , \  \int_\R f(v) v^4 {\rm d} v < \infty \ ,  \ f \in L^\infty \ ,
$$
and set $\mu({\rm d}v) = f(v) {\rm d} v$. Then $\{ [\mu^{\otimes N}]_{\Sp^{N-1}(\sqrt N)} \}_{N \in \N}$
is entropically $\mu$--chaotic, in fact for any $k \in \N$ we have
$$
\lim_{N \to \infty} H\left( M_k ([\mu^{\otimes N}]_{\Sp^{N-1}(\sqrt N)}  ) |\mu^{\otimes k} \right)  = 0
$$

\noindent
Moreover, let $\{ \mu^{(N)}\}_{N \in \N}$ be any family of symmetric probability measures on
$\Sp^{N-1}(\sqrt N)$ with
$$
\frac{1}{N} H\left(\mu^{(N)}|[\mu^{\otimes N}]_{\Sp^{N-1}(\sqrt N)} \right) \to 0 \ .
$$
Then $\{ \mu^{(N)}\}_{N \in \N}$ is entropically $\mu$--chaotic.
\end{theorem}
For a proof the reader may consult \cite{MR2020418}.

It would be nice to know whether or not it is true that for any entropically chaotic family  $\{ \mu^{(N)}\}_{N \in \N}$ for any $k$
$$
\lim_{N \to \infty} H\left( M_k ( \mu^{(N)}  ) |\mu^{\otimes k} \right)  = 0 \ .
$$
This is an open problem. Using a diagonal argument together with the previous theorem one obtains
\begin{corollary}
Let $f$ be a probability density on $\R$ with 
$$
\int_\R f(v) v^2 {\rm d} v = 1 \ , H(f|\gamma) < \infty \ .
$$
Then there exists an $f(v) {\rm d} v$ --entropically chaotic sequence.
\end{corollary}

While exponential entropic decay in time for general initial conditions is false one may ask for
natural conditions on $F_N$ such that
$$
\frac{\langle \mathcal{L}_N, \log F_N\rangle}{ H(F_N|\sigma^{(N)})} \ge c > 0
$$
for some $c$ independent on $N$. This problem is completely open. One could weaken the above question by asking
whether there are  there natural conditions on $f$  so that one can construct a $f$--chaotic sequence $\{F_N\}$ with
$$
\frac{\langle \mathcal{L}_N, \log F_N\rangle}{ H(F_N|\sigma^{(N)})} \ge c > 0
$$
for some $c$ independent on $N$? Nothing is known about this problem either.

\end{document}